\documentclass[prl,twocolumn,superscriptaddress,showpacs,floatfix,linenumbers,footinbib]{revtex4}
\usepackage{graphics}
\usepackage{graphicx}

\begin{document}

\title{Extended interface states enhance valley splitting in
Si/SiO$_2$}
\author{A. L. Saraiva}

\affiliation{Universidade Federal do Rio de
Janeiro, Caixa Postal 68528, 21941-972, Rio de Janeiro, Brazil}

\affiliation{University of Wisconsin-Madison, Madison, Wisconsin 53706, USA}

\author{Belita Koiller}

\affiliation{Universidade Federal do Rio de
Janeiro, Caixa Postal 68528, 21941-972, Rio de Janeiro, Brazil}

\author{Mark Friesen}

\affiliation{University of Wisconsin-Madison, Madison, Wisconsin 53706, USA}

\date{\today}

\begin{abstract}
Interface disorder and its effect on the valley degeneracy of the conduction band edge remains among the greatest 
theoretical challenges for understanding the operation of spin qubits in silicon.
Here, we investigate a counterintuitive effect occurring at Si/SiO$_2$ interfaces.
By applying tight binding methods, we show that intrinsic interface states can hybridize with 
conventional valley states, leading to a large ground state energy gap.
The effects of hybridization have not previously been explored in details for valley splitting.  We find that
valley splitting is \emph{enhanced} in the presence of disordered chemical bonds, in agreement with recent experiments.
\end{abstract}
\maketitle
\pagebreak

\emph{Introduction.}---The transistor revolution has granted silicon heterostructures a special status amongst materials platforms.
Yet after many years of intense study, this system still reveals new and intriguing features.
This is due in part to technological advances that open the door to new physical regimes.  
However, the interest in Si has also been stirred by its unusual materials properties.
The Si conduction band (CB) possesses
six degenerate minima in the first Brillouin zone, known as valleys.
Quantum well confinement and/or the application of uniaxial strain (\textit{e.g.}, in the case of Si/SiGe heterostructures) in the
[001] direction reduces the bulk, cubic symmetry, and raises the energy levels associated with
the transverse $x$ and $y$ valleys~\cite{ando82}. 
At very low temperatures, the CB physics is therefore
governed by the spin and the $z$~valley degrees of freedom.
Control over valley degeneracy is a key concern for Si spin qubits~\cite{koiller01,friesen03,rahman09}.

Experimental~\cite{ando82,takashina06,goswami07} and 
theoretical~\cite{sham79,ando82,saraiva10}
investigations of the physical mechanisms of valley coupling reveal that a
sharp interface between a quantum well and a quantum barrier 
(most commonly SiGe or SiO$_2$) can produce a sizable energy 
splitting between the valley states, and that roughness can suppress this effect~\cite{friesen06,saraiva09}.
Realistic theoretical estimates for the interface-induced valley splitting are on the order
of 0.1-1~meV~\cite{boykin041}, in agreement with many experiments.
However, they cannot explain the recent puzzling results of 
Takashina \textit{et al.}~\cite{takashina06}.  
In an asymmetrically grown Si/SiO$_2$ quantum well, they observe a large ground state gap of 23~meV at the buried oxide (BOX) barrier, but a more typical valley splitting at the second, thermally grown oxide barrier~\cite{takashina04}.

In this work, we demonstrate that conventional CB electron states tend to hybridize with intrinsic Si/SiO$_2$ interface states (IS), which form in the gap.
Hybridization can produce a conducting
ground state that is nondegenerate, due to strong valley orbit coupling. 
The resulting ground state gap can be tens of meV
larger than the valley splitting between pure CB states, which could explain the large values measured in Ref.~\onlinecite{takashina06}. 
Such a gap would be ideal for controlling spin qubits in Si.

\begin{figure}[h!]
\resizebox{70mm}{!}{\includegraphics{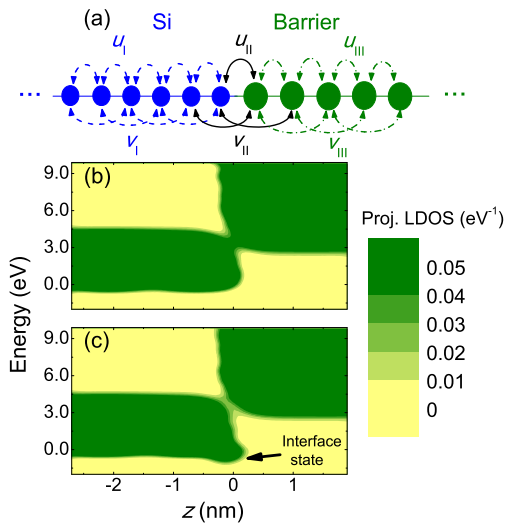}}
\caption{(Color online)
(a) TB model parameters along $z$ for a Si/SiO$_2$ interface.
(b) The projected local density of states (pLDOS), $n^{\text proj}$, as a function of vertical position and energy.
The interface hopping parameters are taken to be $u_{\rm II}=u_{\rm Si}$ and $v_{\rm II}=v_{\rm Si}$.
(c)  The pLDOS obtained for hopping parameters $u_{\rm II}=u_{{\rm SiO}_2}$ and $v_{\rm II}=v_{{\rm SiO}_2}$.  An IS appears in the gap, as indicated by a dip in the CB.  For panels (b) and (c), the results are obtained for a 2D system where model~(ii) is adopted in region~III.}
\label{fig:ldos}
\end{figure}

\emph{Tight Binding Model.}---Since IS are linked to the atomic details of the interface, their theoretical description should likewise be atomistic. 
Here  we study the formation of IS within a minimal, single-electron, two-band tight
binding (TB) model~\cite{boykin041} designed to capture the low-energy physics of the CB.
We consider both one-dimensional (1D) and two-dimensional (2D) versions of the model.
In the 1D version, we consider only hopping along the $z$ direction.  We extend the 1D model of Ref.~\onlinecite{boykin041} to describe the SiO$_2$ barrier as an effective linear chain, as sketched in Fig.~\ref{fig:ldos}(a).
The full chain is composed of three regions:  region~I, with parameters corresponding to bulk Si; region II, involving two sites on either side of the interface; and region~III, with parameters mimicking the SiO$_2$ barrier. 
For the 2D model, we include additional, lateral hopping terms, arranged on a square lattice~\footnote{An accurate description of the geometry of the first Brillouin zone is not a priority here, since transverse Umklapp processes are ruled out at a [001] interface~\cite{saraiva09,saraiva10}.}.
To probe the interface physics, we solve the full TB hamiltonian and we analyze the resulting low energy spectrum and eigenstates.
Details about the models and solutions are presented below.

In region~I, the nearest and next-nearest neighbor vertical ($z$ direction) hopping parameters ($u_{\rm I}$ and 
$v_{\rm I}$) are chosen to reproduce the essential features of the bottom of the CB:  
the Si longitudinal effective mass, $m_l=0.916\, m_0$, and the CB minima, $k_0=\pm 0.82(2\pi/a_{\rm Si})$,
where $a_{\text Si}$ is the length of the Si cubic unit cell.
This gives $u_{\rm I}=u_{\rm Si}=0.68$ eV and $v_{\rm I}=v_{\rm Si}=0.61$ eV. For the 1D model, 
we adopt the onsite energy 
$\epsilon_{\rm I}=\epsilon_{\rm Si} =1.41$~eV, so that the valley minima occur at
zero energy.  We also add an electrostatic potential, $-eFz$, to the onsite parameter, where typical, 
experimental electric fields fall in the range $0.01< F < 0.1$~V/nm.
For the 2D model, we introduce an additional nearest neighbor lateral hopping parameter,
$u^y_I=u^y_{\rm Si}=-10.91$~eV, which gives the correct transverse effective mass $m_t=0.191\, m_0$.
We also modify the onsite parameter  
$\epsilon_{\rm I}=\epsilon^{\rm 2D}_{\rm Si} =23.23$~eV to correctly set the valley minima to zero.

In an empirical TB model, we determine the hopping parameters by matching bulk materials properties.  
The materials properties of the interface are not well known, however.
At a fundamental level, the atomic 
orbitals are locally modified by the inhomogeneous chemical environment~\cite{laughlin80}. 
Such variability is amplified at a Si/SiO$_2$ interface, where important
charge transfer occurs along the Si-O bonds~\cite{laughlin80} and the interface is highly disordered~\cite{helms94}.  
Thus, while region~II is key to understanding IS, the TB hopping parameters are not known with certainty.
To address this problem, we consider a thorough, continuous range of parameters $u_{\rm II}$ and $v_{\rm II}$, as described below.

Although region~III is better understood than region~II, 
TB models of SiO$_2$ are still not well established.
We therefore explore three
minimal models for region~III, to observe how new behaviors emerge, not aiming for a quantitative description. Our results do not depend qualitatively on the particular oxide model, as will be shown. The three
models we consider are given as follows.

(i) An indirect gap crystalline material.  In this case, we use the same hopping parameters as Si, so that $u_{\rm III}=u_{\rm I}$, $v_{\rm III}=v_{\rm I}$, and $u^y_{\rm III}=u^y_{\rm I}$.  Only the band edge is shifted upward, with the experimental offset of $W\approx 3.0$~eV.  The onsite parameter is then given by $\epsilon_{\rm III}=\epsilon_{\rm I}+W$.

(ii) A direct gap crystalline material.  In this case, there is a single valley at $k=0$, and we adopt  the effective, isotropic mass of $0.34\, m_0$, corresponding to FD$_3$M $\beta$-cristobalite~\cite{ramos04}.  This is achieved by setting $u_{\rm III}=u_{{\rm SiO}_{2}}=3.28$~eV and  $v_{\rm III}=0$.  The shifted band edge is set by $\epsilon_{\rm III}=\epsilon_{{\rm SiO}_2}+ W=9.56$~eV.

(iii) An amorphous material.  There is strong evidence that this is the most accurate model scenario~\cite{helms94}, although crystalline phases of SiO$_2$ have been reported within a few layers of the interface~\cite{yamasaki01}, with direct gap $\beta$-cristobalite being the most abundant phase.  Our amorphous model consists of TB parameters chosen randomly in the range
$u_{\rm III}\in[u_{{\rm SiO}_{2}}-0.5,u_{{\rm SiO}_{2}}+0.5]$ and 
$v_{\rm III}\in[-0.5,0.5]$, in units of eV. For simplicity we take 
the onsite energies and transverse hoppings to be the same as model (ii). Thus, the lateral disorder
is mediated entirely by the vertical hopping parameters.

\emph{Interface States.}---Intrinsic IS may occur even at smooth, ordered interfaces.
This is in contrast with extrinsic IS, which require broken bonds, impurities, or some type of disorder potential.
The intrinsic IS are often referred to as Tamm/Shockley states~\cite{tamm32,shockley39}. 
We now demonstrate how they can emerge naturally from our Si/SiO$_2$ TB models. 
Later, we will study the dependence of IS on specific interface models. 

The electron affinity varies abruptly at a sharp interface, corresponding to a sudden change of the onsite terms in our TB model.  On the other hand, chemical bonds are modified by their local environment, leading to a more gradual variation of the hopping terms over several monolayers. The energy of the lowest orbital level therefore varies locally.  This is the origin of the IS \footnote{Note that IS are not expected to occur in Si/SiGe because both semiconductors have a similar CB structure and the interface is well ordered (\textit{i.e.}, epitaxial).}.  

To demonstrate the emergence of IS in our TB model, we first compute the local density of states (LDOS), defined as $n(i,j,E)= -\frac{1}{\pi}{\rm Im}[G(i,j;i,j;E)]$, for the coordinate indices $(i,j)$. The Green's function matrix $G(i,j;i^\prime,j^\prime;E)$ is obtained from $G(E)=1/(E-H+i0)$.  In the 2D case, we then compute the projected LDOS (pLDOS) along $z$, defined as $n^{\rm proj} (j,E) = \sum_i {n(i,j,E)}$. 

The pLDOS provides a means to visualize the local CB, as shown in Figs.~\ref{fig:ldos}(b) and (c).  For demonstration purposes, we consider two cases.
In Fig.~\ref{fig:ldos}(b), we take  $u_{\rm II} = u_{\rm Si}$ and $v_{\rm II} = v_{\rm Si}$. The CB appears as a dark shaded region, with a simple step in the band edge at the interface, $z=0$. 
In Fig.~\ref{fig:ldos}(c), we take $u_{\rm II}=u_{{\rm SiO}_{2}}$ and $v_{\rm II}=v_{{\rm SiO}_{2}}$.  In this case, the lower edge of the CB dips by over 200~meV in region~II.  The dip signifies an IS, which is strongly
peaked at the interface, and decays over a few \AA\, on either side.
The localized nature of the wavefunction suggests that the computed
IS is sensitive to the values used for $u_{\rm II}$ and $v_{\rm II}$.

\begin{figure}[h!]
\resizebox{70mm}{!}{\includegraphics{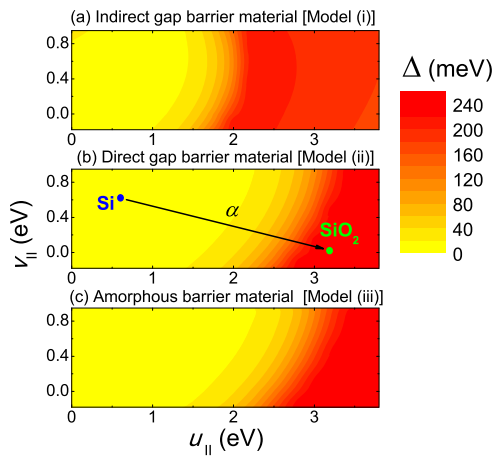}}
\caption{(Color online)
Ground state gap $\Delta$ as a function of the
TB hopping parameters in region~II.  Results for the three barrier models (i)-(iii) are presented in panels (a)-(c) respectively.
In each case, we assume an electric field $F=0.05$~V/nm, and a 1D supercell that is large enough to avoid finite size effects.
The circles in panel (b) indicate asymptotic points in region~II, where the hopping parameters correspond to Si or SiO$_2$.
The parameter $\alpha\in [0,1]$ interpolates linearly between these points. }
\label{fig:sweep}
\end{figure}

\emph{Ground state gap.}---Fig.~\ref{fig:sweep} shows the calculated ground state energy gap $\Delta$, as a function of the interface hopping parameters $u_{\rm II}$ and $v_{\rm II}$. 
Results are shown for each barrier model, (i)-(iii), in region~III.
In each case, we note that $\Delta$ is determined principally by the hopping parameter $u_{\rm II}$, rather than $v_{\rm II}$.  
For small $u_{\rm II}$, $\Delta \simeq 7$~meV corresponds to the conventional valley splitting, induced by the sharp interface.  
For large $u_{\rm II}$, a much larger gap emerges, $\Delta \simeq 200$~meV, which cannot be explained by valley splitting, indicating an IS, as discussed below.

Comparison between the three panels of Fig.~\ref{fig:sweep} reveals that the TB model used in region~III does not qualitatively affect the nature of the IS.
This is because the step in the CB is so large that the electron barely penetrates into region~III.  
For this reason, and for simplicity, we adopt model~(ii) throughout the remainder of the paper.

Since the gap $\Delta$ does not exhibit strong features in Fig.~\ref{fig:sweep}, we consider two representative special cases, ($u_{\rm II}=u_{\rm Si},\, v_{\rm II} = v_{\rm Si}$)
and ($u_{\rm II} = u_{\rm SiO_2},\, v_{\rm II} = v_{\rm SiO_2}$), as indicated by circles in 
Fig.~\ref{fig:sweep}(b), and the line connecting these points. This line can be expressed parametrically through the equations
$u_{\rm II}=(1-\alpha) u_{\rm Si} + \alpha\, u_{{\rm SiO}_2}$ 
and $v_{\rm II}=(1-\alpha)v_{\rm Si} + \alpha\, v_{{\rm SiO}_2}$.  We may then consider just one parameter,
$\alpha\in [0,1]$, which interpolates between a conventional CB state ($\alpha=0$) and an IS state ($\alpha=1$).

\emph{Hybridization of IS and CB States.}---We now demonstrate how IS can emerge and compete with valley states through hybridization, leading to large ground state gaps in the right-hand 
regions of Fig.~\ref{fig:sweep}.

\begin{figure}[h!]
\resizebox{80mm}{!}{\includegraphics{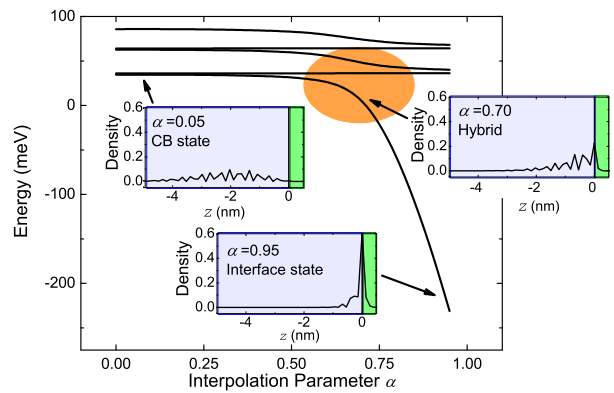}}
\caption{(Color online) The five lowest energy levels in the 1D TB model, for an electric field of 0.01 V/nm. 
The insets show the ground state electron probability densities corresponding to the representative values $\alpha=0.05,$ 0.70, and 0.95. The region where strong hybridization occurs is highlighted.}
\label{fig:alpha}
\end{figure}

The five lowest energy eigenvalues of the 1D TB hamiltonian are shown as a function of the interpolation parameter $\alpha$ in Fig.~\ref{fig:alpha}~\footnote{Note that a single, discrete IS is obtained from the 1D model.  This is in contrast with the continuum of levels observed in Fig.~\ref{fig:ldos}, where we used a 2D model.}.
On the left-hand side, we observe nearly degenerate doublets.  We refer to these as valley pairs because they share an envelope, but their fast oscillations are $\pi/2$ out of phase.  
In this regime, it is appropriate to equate the ground state gap with the valley splitting $\Delta_V$.  Its value of $\sim 2$~meV is smaller than Fig.~\ref{fig:sweep} because the electric field $F=0.01$~V/nm is lower.

On the right-hand side of Fig.~\ref{fig:alpha}, an IS splits off from the CB into the band gap. In this regime, the ground state gap becomes quite large. However, the wavefunctions of the ground and excited states are very different, and they do not share an envelope, so it is not appropriate to speak of valley splitting.

At intermediate values of $\alpha\simeq 0.5-0.8$ the IS hybridizes with the CB states, and we observe characteristic level anti-crossings, highlighted in Fig.~\ref{fig:alpha}. The mixing involves only the lowest state in each doublet, indicating a valley selection rule. Such IS hybridization always enhances the ground state gap compared to $\Delta_V$.

The insets show the electronic probability density of the lowest 1D eigenstate, obtained for three values of $\alpha$.  We observe characteristic CB and IS features, consistent with our previous discussion.  The intermediate, hybrid wavefunction exhibits characteristics of both functions.

\emph{Disordered Interfaces.}---The hybridization of IS and CB states depends strongly on the unknown and locally varying hopping
parameters of region II. The disorder can occur laterally, so that conditions favoring CB states or IS may coexist in a given heterostructure.
Microscopically, the variations in hopping parameters can be interpreted as the stretching and bending of Si-O bonds to accommodate crystalline imperfections and the natural buckling of the interface.  In some cases, an O atom may be completely absent, causing a missing link for charge transfer, which we model here as $u_{II}=0$.
We now show
that lateral disorder in the TB parameters may actually enhance the ground state gap, in contrast with interface height disorder, which suppresses the valley splitting~\cite{friesen06}.  We also investigate the possibility of lateral localization of a hybridized IS, in the presence of disorder.

\begin{figure}[t]
\resizebox{70mm}{!}{\includegraphics{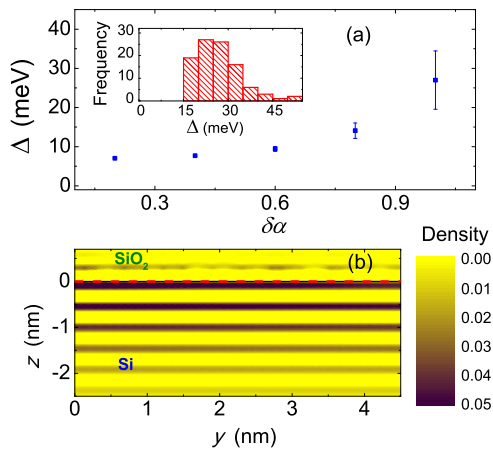}}
\caption{(Color online)
(a) The average ground state gap as a
function of the disorder amplitude $\delta\alpha$, for 100 realizations of a supercell containing $N_y=50$ atomic layers in the $y$ direction (2D calculation). The error bars correspond to a standard deviation of one sigma. The histogram inset shows the distribution of ground state gaps for the case
$\delta\alpha=1$. 
(b) The 2D electronic probability
density $|\Psi(y,z)|^2$ is shown for a ``worst case" disorder realization.
The dashed line indicates the interface position.
All results assume $F=0.05$ V/nm.}
\label{fig:disorder}
\end{figure}

We introduce disorder into our 2D TB model by assigning random interpolation parameters $\alpha$ for each position $y$ along the interface, in the range $[-0.26,\delta\alpha]$, keeping $\delta \alpha$ constant in each realization. The lower bound of this range corresponds to the absent link condition, $u_{\rm II}=0$. 
The upper bound $\delta \alpha$ characterizes the degree of the disorder.
Figure~\ref{fig:disorder}(a) shows the mean ground state gap $\Delta$ as a
function of $\delta\alpha$.
We conclude that any amount of bond disorder that can be modeled in this way will enhance $\Delta$.  
When the interpolation parameter $\alpha$ ranges all the way to $\delta \alpha\simeq 1$,
the resulting gap is consistent with the large experimental value reported
in Ref.~\onlinecite{takashina06}.

The error bars shown in Fig.~\ref{fig:disorder}(a) indicate the standard deviation in values of $\Delta$ for different disorder realizations. In the inset, we show a histogram for the frequency distribution of the ground state gap, for the disorder amplitude $\delta\alpha=1$.
In this case, all realizations give $\Delta> 15$ meV, indicating that the bond disorder consistently enhances the 
ground state gap by a large amount.

Large energy variations, caused by lateral disorder, potentially induce IS localization.  Such localized states might not be 
visible in transport measurements like those of Ref.~\onlinecite{takashina06}.  A detailed study of localization is beyond the scope of this work.  However, we can check for lateral localization over smaller length scales.
In Fig.~\ref{fig:disorder}(b) we plot the 
2D electronic probability density for a ``worst case" disorder realization, where the ground state gap is particularly large, with $\Delta=36$~meV. 
We observe that the wavefunction spreads out uniformly across the lateral plane, due to its vertical extension far outside of region~II where the disorder occurs.  For a high quality Si/SiO$_2$ interface of width $\sim 1$~\AA~\cite{ando82}, this suggests that the vertical extension of the hybridized state may be able to overcome localization effects.  This is in contrast with extrinsic IS, which are typically localized. Note that disorder in the lateral hopping parameters, not included here, may also contribute to localization.

\emph{Conclusions.}---We have shown that a typical ground state
of an electron confined at a Si/SiO$_2$ interface is a
hybrid of interface and CB states. The IS component causes a sizeable ground state gap, while the CB component is vertically extended, providing conduction characteristics similar to a pure CB state. Hybridization therefore provides a plausible explanation for the measurements reported in Ref.~\onlinecite{takashina06}, without invoking many-body physics.
We also conclude that bond disorder may increase the ground state gap by enhancing hybridization.  At Si/SiO$_2$ interfaces, this mechanism competes with the more widely studied suppression of valley splitting, which occurs at a rough interface~\cite{friesen06}. The net enhancement or suppression of valley splitting then depends upon the details of the heterostructure.

Our results suggest that it may be possible to tune the ground state gap and the localization properties of a Si/SiO$_2$ interface by controlling the hybridization of IS and CB states in the ground state wavefunction. 
Control parameters could include electric or magnetic fields, providing new directions for theoretical and experimental investigations. More detailed studies of atomic scale Si/SiO$_2$ bond structures and realistic disorder models are also needed, to help pinpoint how the fabrication process may be used to engineer the ground state hybridization.

\begin{acknowledgments}
We acknowledge X. Hu and S. Coppersmith for many fruitful discussions. This work was supported in part by ARO and
LPS, by NSF and CAPES. BK thanks CNPq, FUJB, INCT on Quantum Information and FAPERJ.

\end{acknowledgments}
\bibliography{is-long}

\end{document}